\documentclass[doublecol]{epl2} 
\def\etcl{$\kappa$-(BE\-DT\--TTF)$_2$\-Cu\-[N\-(CN)$_{2}$]Cl}

\def\etcn{$\kappa$-(BE\-DT\--TTF)$_2$\-Cu$_2$(CN)$_{3}$}
\def\cm{cm$^{-1}$}
\def\etal{{\it et al.}}

\usepackage{graphicx}

\title{Search for quantum electronic dipoles in the
dimerized $\kappa$-(BE\-DT\--TTF)$_2$\-Cu\-[N\-(CN)$_{2}$]Cl salt}
\shorttitle{Search for quantum electronic dipoles in the
$\kappa$-ET$_2$\-Cu\-[N\-(CN)$_{2}$]Cl dimerized salt}

\author{
     S. Tomi\'{c}\inst{1}
\and M. Pinteri\'{c}\inst{1,2}
\and T. Ivek\inst{1,3}
\and K. Sedlmeier\inst{3}
\and R. Beyer\inst{3}
\and D. Wu\inst{3}
\and J. A. Schlueter\inst{4}
\and D. Schweitzer\inst{3}
\and M. Dressel\inst{3}
}

\shortauthor{S. Tomi\'{c} \etal}

\institute{                    
  \inst{1} Institut za fiziku - P.O.\ Box 304, HR-10001 Zagreb, Croatia\\
  \inst{2} Faculty of Civil Engineering, University of Maribor - Smetanova 17, 2000 Maribor, Slovenia\\
  \inst{3} Physikalisches Institut, Universit\"{a}t
Stuttgart - Pfaffenwaldring 57, D-70550 Stuttgart Germany\\
  \inst{4} Material Science Division, Argonne National Laboratory - Argonne, Illinois 60439-4831, U.S.A.
}

\pacs{71.30.+h}{Metal-insulator transitions and other electronic transitions}
\pacs{77.22.Gm}{Dielectric loss and relaxation}
\pacs{78.30.Jw}{Infrared spectra of organic compounds, polymers}
\abstract{
The Mott insulator $\kappa$-(BE\-DT\--TTF)$_2$\-Cu\-[N\-(CN)$_{2}$]Cl consists
of molecular dimers arranged on an anisotropic triangular lattice and develops a
canted antiferromagnetic ground state. It has recently been suggested that this
system features purely electronic ferroelectricity which requires an electric
dipole moment. Optical spectroscopy clearly rules out charge imbalance in this
system, which excludes the existence of quantum electric dipoles on the dimers
and subsequently a dipolar spin coupling. We suggest that the prominent in-plane
dielectric response in $\kappa$-(BE\-DT\--TTF)$_2$\-Cu\-[N\-(CN)$_{2}$]Cl is due
to short-range discommensurations of the antiferromagnetic phase in the
temperature range $30 < T < 50$~K, and domain wall relaxations at lower temperatures.}

\begin{document}

\maketitle

\section{Introduction}
The interplay of electronic correlations and magnetic frustration causes various
exotic ground states, such as Mott insulator and quantum spin liquid, that have
drawn enormous attention in recent years \cite{Lee08,Balents10}. Examples are
exotic quantum phases on kagome and honeycomb lattices explored in graphene and
transition metal compounds, for instance
\cite{Kitaev06,Meng10,Varney11,Nakatsuji12}. The realization by organic
conductors with triangular lattices turns out to be especially suitable because
they can be nicely tuned by slight chemical and physical variations
\cite{Kanoda11,Powell11}. The $\kappa$-(BEDT-TTF)$_2X$ materials serve as prime
examples in this regard, spanning from the Fermi-liquid metal $X$ = Cu[N(CN)$_2$]Br
that superconducts below 12~K, to the first example of a spin-liquid system
found in $X$ = Cu$_2$(CN)$_3$ despite the strong exchange interaction of
$J=250$~K within the triangular lattice \cite{Shimizu03,Kurosaki05}. In this
work we take a closer look at the Mott insulator $X$ = Cu[N(CN)$_2$]Cl which
shows canted-spin antiferromagnetic ordering, {\it i.e.}, weak ferromagnetism at
temperatures below 30~K
\cite{Welp92,Kawamoto95,Pinteric99,Smith2003}.\footnote{Although first
experiments showed antiferromagnetism below 45~K, all subsequent measurements
revealed the antiferromagnetic ordering and canting in the temperature range
20~K--30~K.}

\begin{figure}
\centering\includegraphics[clip,width=0.8\columnwidth]{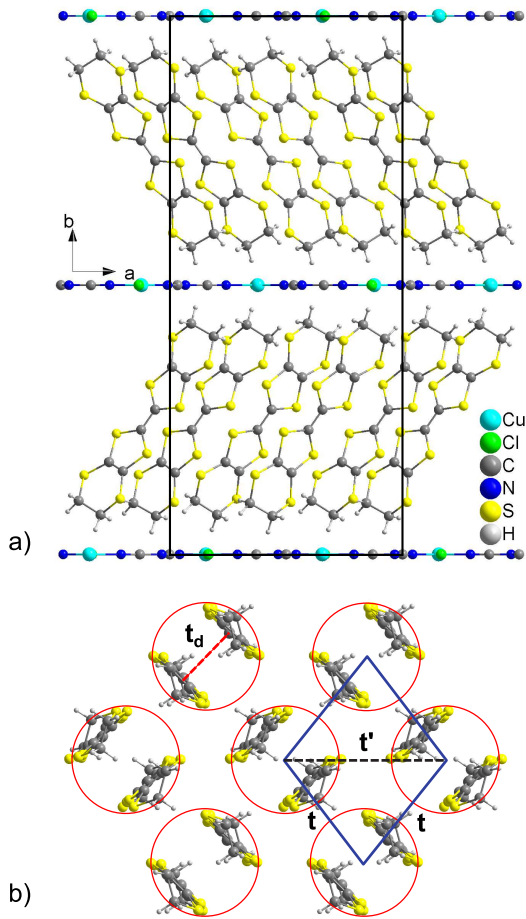}
\caption{(Color online) a) The side view on \etcl{} demonstrates the staggered
layers of BEDT-TTF molecules separated by sheets of polymeric anions.
b) For $\kappa$-(BEDT-TTF)$_2X$ in general, the molecules are arranged in dimers which
constitute an anisotropic triangular lattice within the plane. The interdimer
transfer integrals are labeled by $t$ and $t^{\prime}$, and intradimer by
$t_d$.}
\label{fig:structure}
\end{figure}

The structure of $\kappa$-phase materials based on the
bis\--(e\-thy\-le\-ne\-di\-thio)\-te\-tra\--thi\-a\-ful\-va\-le\-ne (BEDT-TTF)
molecule consists of two-dimensional layers of dimers arranged in an anisotropic
triangular lattice. In particular, \etcl{} forms an orthorhombic two-layer
system (anion layers in $ac$-plane, $b$-axis perpendicular to it) with four
BEDT-TTF dimers per unit cell due to the tilting along the $a$-axis in opposite
direction for adjacent layers, as displayed in Fig.\ \ref{fig:structure} a)
\cite{Williams90}. Conducting layers of cationic BEDT-TTF molecules are
separated by essentially insulating anion sheets. The BEDT-TTF donors form
face-to-face dimers which themselves are rotated by about 90$^\circ$ with
respect to neighboring dimers, as sketched in Fig.\ \ref{fig:structure} b).

With an average of half a hole per molecule, the common theoretical description
of $\kappa$-structures considers each dimer as a single site (an effective
half-filled band) and applies a Hubbard model with strong on-site Coulomb
repulsion $U$. Although it has captured most of the physics
\cite{Kino95,Kanoda97}, this approach was put into question recently because
dielectric measurements yield certain features in \etcn{}
\cite{AbdelJawad10,Poirier12} and \etcl{} \cite{Lunkenheimer12} that are
interpreted as ferroelectric response due to charge disproportionation within
the dimers. Consequently Hotta proposed a dipolar-spin liquid \cite{Hotta10}
taking the quantum electric dipoles on the dimers into account that interact
with each other through the dipolar spin coupling resulting in a quarter-filled
two-band Hubbard model with intersite Coulomb interaction $V$. Similar
considerations have been put forward by other groups \cite{Gomi10,Li11}.
Notably, a recent {\it ab initio} study of $\kappa$-phase materials applying a
single-band extended Hubbard model \cite{Shinaoka12} discovered that charge
fluctuations are enhanced by the inter-dimer Coulomb interaction. Moreover,
Tocchio \etal{}\ \cite{Tocchio2009} showed that the magnetic phase can stabilize
in presence of charge fluctuations. Here we present our dielectric and optical
experiments on the dimerized \etcl{} that are at odds with presented models. Our
results call for reconsideration of the coupling between charge and spin degrees
of freedom.

\section{Experimental and Results}
Single crystals of the quasi-two-dimensional organic charge-transfer salt
\etcl{} (herafter $\kappa$-Cl) were grown by standard electrochemical methods
\cite{Williams92}.

We revisit our ac measurements of complex conductivity on $\kappa$-Cl within the
molecular planes $\mathbf{E}\parallel ac$ \cite{Pinteric99}. Complex dielectric
spectra were obtained at temperatures 10--50~K by employing an LCR meter
in the range 20~Hz--1~MHz. Two important improvements in data analysis have been
implemented. First, stray-impedance background contributions of the sample
holder have been measured in the open circuit configuration and subtracted from
the sample measurements.
To the resulting complex dielectric spectra $\varepsilon_\parallel(\omega)$ a sum of two
generalized Debye functions was fitted, 
$ \Delta\varepsilon_{\parallel,1} / [ 1 + \left(i \omega \tau_{0,1} \right)^{ 1-\alpha_1} ]
+ \Delta\varepsilon_{\parallel,2} / [ 1 + \left(i \omega \tau_{0,2} \right)^{ 1-\alpha_2} ]
+ \varepsilon_{\parallel,\mathrm{HF}}$,
where $\Delta\varepsilon_\parallel$ is the dielectric strength, $\tau_0$ the
mean relaxation time, $1-\alpha$ the symmetric broadening of the relaxation time
distribution function of the two dielectric relaxation modes, and
$\varepsilon_{\parallel,\mathrm{HF}}$ is the high-frequency dielectric constant. In
particular, this approach successfully resolves the modes when they are near
boundaries of the experimental frequency window. The second improvement extends
the studied temperature range: above 30~K we determine the dielectric strength
from capacitance measured between $10^5$ and $10^6$~Hz (well below the
dielectric relaxation), which enables us to track the dielectric response up to
50~K. In the remainder of this paper we discuss in detail only the results
obtained for Sample 2 of Ref.\ \cite{Pinteric99}, which we consider
quantitatively more reliable due to a negligible stray impedance after
subtraction of background. An influence of other extrinsic effects, such as
contact resistances and surface layer capacitances, was ruled out following the
strict procedure explained in Ref.\ \cite{Ivek08}.

\begin{figure}
\centering\includegraphics[clip,width=0.8\columnwidth]{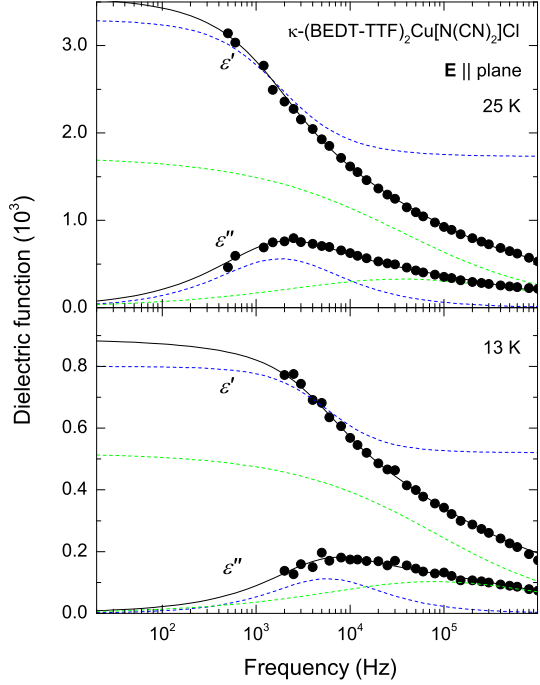}
\caption{(Color online) Real and imaginary parts of the in-plane dielectric function of $\kappa$-Cl
at two temperatures ($T=13$~K and 25~K). Full lines are fits to the sum
of two modes, given by the green and blue dashed lines.}
\label{fig:dsmodes}
\end{figure}

\begin{figure}
\centering\includegraphics[clip,width=0.8\columnwidth]{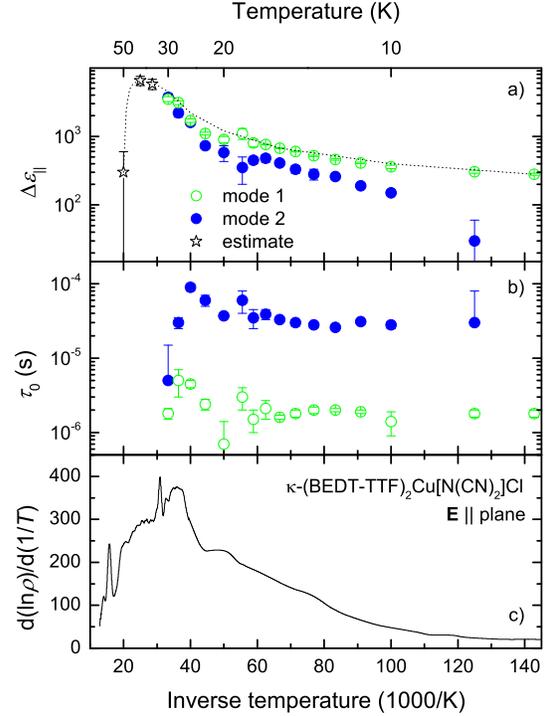}
\caption{(Color online) a) In-plane dielectric strength and b) mean relaxation time of
$\kappa$-Cl as a function of inverse temperature. Empty green and full blue circles
represent parameters of the mode 1 and mode 2, respectively. Above 30~K both
modes are outside of the experimental frequency window and only the total
dielectric strength $\Delta\varepsilon_{\parallel}$ can be extracted, denoted by
stars. Dotted line is guide for the eye. c) The in-plane resistivity derivative
shows a broad maximum which indicates the formation of a low-temperature phase
at short-range scale. Dielectric response emerges in the region of resistivity
crossover.}
\label{fig:dsparams}
\end{figure}

If the charge is unequally distributed among the molecules of the dimers, we may
envision an electronic dipole moment within the layers. From our ac measurements
on $\kappa$-Cl in the $ac$-plane we in fact see a very strong dielectric
relaxational response [measured by the difference between static and
high-frequency dielectric constant $\Delta\varepsilon_{\parallel}=
\varepsilon_\parallel(0) - \varepsilon_{\parallel,\mathrm{HF}} \approx 3000$]
that emerges as the temperature drops below 50~K. The asymmetric shape of the
dielectric function plotted in Fig.\ \ref{fig:dsmodes} clearly indicates the
presence of two modes in the kHz range contributing with comparable strength.
From the fits of $\varepsilon_{\parallel}^{\prime}(\omega)$ and
$\varepsilon_{\parallel}^{\prime\prime}(\omega)$ at a certain temperature we can
extract their dielectric strength and mean relaxation time, shown in
Figs.\ \ref{fig:dsparams} a), b) as a function of $1/T$. The large error bars at 50~K
reflect the difficulty in determining the modes at this temperature. At very low
temperatures the dielectric strength of the first mode saturates at finite
values of about $\Delta\varepsilon_{\parallel}\approx 300$, while the dielectric
strength of mode 2 drops rapidly and eventually disappears at 7~K. The
dielectric data on $\kappa$-Cl show that neither mode moves with temperature.
The broadening parameter $1-\alpha$ of modes 1 and 2 is $0.45\pm 0.05$ and $0.80
\pm 0.10$, respectively.

Recently Lunkenheimer \etal{}\ \cite{Lunkenheimer12} measured the electric
polarization switching and also the dielectric constant of $\kappa$-Cl
perpendicular and parallel to the BEDT-TTF planes, the magnitude of latter
comparable to our results. When the temperature is reduced, the perpendicular
dielectric constant peaks around 25~K. The maximum value increases for low
frequencies up to $\varepsilon_{\perp}^{\prime}(\nu=2.1~{\rm Hz}) \approx 420$.
Since the dielectric anomaly occurs around the same temperature as magnetic
ordering, they claim a loss of spin frustration due to ferroelectric ordering,
resulting in antiferromagnetic spin order. However, a clear link of the magnetic
order to the dielectric anomaly could not be presented. Most importantly, there
is no experimental evidence of charge disproportionation on the dimers which is
crucial for the ferroelectric interpretation.

\begin{figure}
\centering\includegraphics[clip,width=0.8\columnwidth]{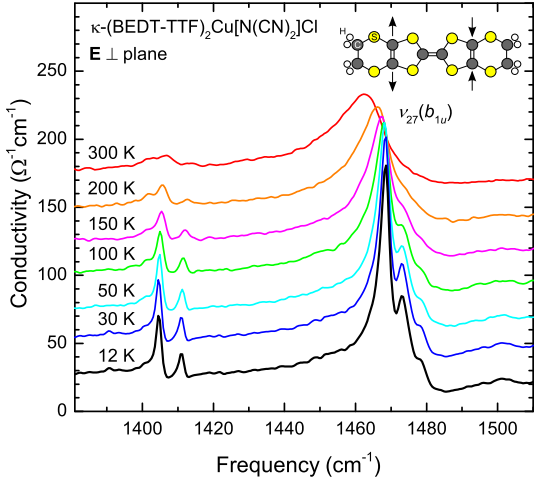}
\caption{(Color online) Temperature evolution of the out-of-plane
optical conductivity of \etcl{} measured at the narrow side of crystals.
For clarity reasons the curves are shifted with respect to each other.
Arrows on the sketched BEDT-TTF molecule indicate the antisymmetric C=C
stretching vibrational mode $\nu_{27}(b_{1u})$ which is a very sensitive
local probe of charge per molecule.}
\label{fig:IR}
\end{figure}

In order to clarify this point, we have performed comprehensive in- and
out-of-plane infrared measurements on $\kappa$-Cl single crystals down to 12~K.
The frequencies of certain intramolecular vibration modes in BEDT-TTF crystals
strongly depend on molecular charge, which makes Raman and infrared spectroscopy
the superior local probe for the investigation of charge distribution
\cite{Maksimuk01,YamamotoGirlando,Sedlmeier12}. Particular emphasis was put on
the most charge-sensitive intramolecular vibrational modes
$\nu_{2}(a_{\mathrm{g}})$, $\nu_{3}(a_{\mathrm{g}})$, and
$\nu_{27}(b_{1\mathrm{u}})$. The last involves the antisymmetric stretching
vibration of the outer C=C bond as sketched in Fig.\ \ref{fig:IR} and can only
be observed perpendicular to the crystal plane. Measurements were taken on the
thin sides of the crystals with 1~\cm{} resolution. Kramer-Kronig analysis was
performed using a constant reflectivity extrapolation at low frequencies and
temperatures.

In Fig.\ \ref{fig:IR} the mid-infrared conductivity is plotted for different
temperatures. Neighbouring the $\nu_{28}(b_{1u})$ doublet slightly above
1400~\cm{}, we observe the charge-sensitive $\nu_{27}(b_{1\mathrm{u}})$ mode at
1460~\cm{}, where it is expected for half a hole per BEDT-TTF molecule
\cite{Maksimuk01,YamamotoGirlando}. With decreasing temperature there is a
slight hardening of a few \cm{} and a strong narrowing. In $\kappa$-Cl the
molecules in adjacent layers are tilted in opposite directions as depicted in
Fig.\ \ref{fig:structure} a). The crystallographic inequality among the eight
BEDT-TTF per unit cell causes two satellites to evolve at the high-frequency
wing of the $\nu_{27}(b_{1\mathrm{u}})$ mode.

The important observation is that the molecular vibrational modes do not split
upon cooling, giving no indication of charge redistribution in $\kappa$-Cl as
temperature is reduced. A similar conclusion can be drawn from the electron
molecular-vibrational coupled totally-symmetric modes $\nu_{2}(a_g)$ and
$\nu_{3}(a_g)$ \cite{Sedlmeier12}. According to the strong dependence of the
eigenfrequency $\nu_{27}(b_{1\mathrm{u}})$ on ionicity of $-140$~\cm{} per
electron \cite{Maksimuk01,YamamotoGirlando}, we can exclude a charge imbalance
of more than 1\%. A charge disproportionation of $\delta\rho=0.6e$, typical for
charge-ordered systems such as $\alpha$-(BEDT-TTF)$_2$I$_3$ \cite{Ivek11},
results in a splitting of 80~\cm. Nothing like that is found, neither in our
data \cite{Sedlmeier12}, nor in temperature-dependent Raman and infrared
experiments by other groups \cite{TruongEldridge,Maksimuk01}. $\kappa$-Cl
becomes insulating due to Mott-Hubbard localization of one hole per BEDT-TTF
dimer, rather than the formation of charge order.

\section{Discussion}
The question now arises whether the dielectric relaxation observed in
$\kappa$-Cl in and out of plane can be interpreted as a consequence of
ferroelectricity, and in which way the theoretical model has to be modified. The
ferroelectric approach very successfully describes commensurate charge ordering
in the one-dimensional organic compounds (TM\-TTF)$_2X$, for which charge
disproportionation was established by NMR and optical measurements
\cite{Takahashi06Dressel12}. Ferroelectricity becomes apparent in the divergence
of dielectric constant at the transition temperature, where the two branches of
$1/\Delta\varepsilon_{\parallel}(T)$ above and below $T_{\rm CO}$ are very close
to linear and the slope at $T < T_{\rm CO}$ is two times larger than at
$T > T_{\rm CO}$. In two-dimensional electron systems, such as
$\alpha$-(BEDT\--TTF)$_2$I$_3$ or the ladder compound
(Sr,Ca)$_{14}$\-Cu$_{24}$\-O$_{41}$, the situation is already quite different
\cite{Vuletic06,Ivek08,Ivek11} in that no evidence for ferroelectric character
of the CDW phase transition has been found. Indeed, the in-plane dielectric
response in $\kappa$-Cl does not show this signature of ferroelectricity,
either.

Let us consider the magnitude of $\varepsilon$, both in-plane and perpendicular
to the molecular planes of $\kappa$-Cl. The out-of-plane dielectric response
$\varepsilon_{\perp} \approx 10 -100$ \cite{Lunkenheimer12} appears only for
$T<35$~K and is more than an order of magnitude smaller than
$\varepsilon_{\parallel}$, which is present already below 50~K. These values of
$\varepsilon_{\perp}$ are typical for two-dimensional organic compounds
\cite{Inagaki0304,AbdelJawad10}. Previously, dielectric constants of
$\varepsilon_{\parallel}\approx 10^4$ and larger have been successfully
attributed to the long-wavelength charge-order excitations, while
$\varepsilon_{\parallel}<10^3$ is related to short-wavelength charge excitations
\cite{Pinteric99,Vuletic06,Ivek11}; only the latter seem to be present in the
$\kappa$-Cl system.

The data on $\kappa$-Cl evidence that a dielectric constant suddenly emerges
below 50~K, reaches a maximum at $T_\mathrm{cross} \approx 40$~K and decreases
below. $T_\mathrm{cross}$ corresponds well to the broad maximum as shown by the
in-plane resistivity derivative, Fig.\ \ref{fig:dsparams} c). The appearance of
dielectric response additionally coincides with the onset of the finite charge
gap at 50~K and the enhancement of the antiferromagnetic spin correlations
\cite{Kezsmarki06,Elsaesser12}. Hence, the broad dielectric feature can be taken
as an indication of a crossover into a new low-temperature phase which develops
at short length scales. Nuclear magnetic and electron spin resonance
measurements \cite{Kawamoto95,Kubota96,Miyagawa2004,Yasin11} yield a weak
ferromagnetic state below $T_N \approx 30$~K with substantial short-range
fluctuations extending up to 60~K. Tanatar \etal{}\ proposed the formation of
antiferromagnetic structure at temperatures as high as 70~K based on their
observation of long-term (100~s) resistance relaxations both in Hall effect and
magneto-resistance produced by a rotation of magnetic field \cite{Tanatar97}.
This leads to the picture of magnetic domains which develop below 70~K and cause
the metastable resistance across the domain walls due to soliton formation; the
transient effects are largest at about 30~K and strongly drop below.

Associated with these domain walls, we suggest pairs of charged defects which
respond to ac electric fields and carry spin as well as charge. While at
elevated temperatures these are commensurate fluctuating antiferromagnetic
domains, ferromagnetic domains develop at low temperatures. The dielectric mode
2 can thus be attributed to discommensurations of the otherwise commensurate
antiferromagnetic phase which develops at short length scales below about 50~K.
Once the antiferromagnetic phase with canted spins has formed at long scales
(around $20-30$~K) the dielectric response is taken over by domain walls of this
ferromagnetic phase. The dielectric mode 1 then becomes the dominant
contribution to the dielectric polarization at low temperatures. A similar
conclusion can be drawn from the anisotropy of the dielectric constant that is
considerable in the temperature region of antiferromagnetic fluctuations
($\varepsilon_\parallel/\varepsilon_\perp \geq 100$), but becomes much smaller
($\varepsilon_\parallel/\varepsilon_\perp \approx 5$--10) in the weak
ferromagnetic state \cite{Lunkenheimer12}. This effect happens at short length
scales \cite{Tokura89} but is not associated with macroscopic intra- and/or
intermolecular distortions, as supported by infrared spectroscopy. Our findings
are in line with theoretical consideration by Naka \etal{}\ \cite{Naka10}
showing that magnetic and ferroelectric phases are exclusive to each other. In
the end and in line with infrared results, no indication that points to a charge
disproportionation was found in nuclear magnetic resonance data \cite{Kanodapc}.

\section{Conclusions}
Vibrational spectroscopy rules out any charge disproportionation in the
dimerized \etcl{} system with antiferromagnetic ground state and canted spins.
No appreciable change in the charge-sensitive vibrational features is revealed
upon cooling, providing evidence that the charge distribution is not altered
with decreasing temperature. The absence of electric dipoles on the dimers rules
out ferroelectricity and is in accord with the low dielectric constant compared
to known charge ordered systems \cite{Ivek11}. The in-plane dielectric response
of \etcl{} is rather explained by short-range discommensurations of the
antiferromagnetic phase in the temperature range $30$~K$ < T < 50$~K, and domain
wall relaxations at lower temperatures. The weak dielectric response observed in
the perpendicular direction is caused by the layered structure of these salts.
Further theoretical efforts are needed to clarify the origin of dielectric
relaxation observed in absence of charge disproportionation.

\acknowledgments We thank S. Mazumdar, K. Kanoda, J.-P. Pouget, M. Tanatar and
T. Yamamoto for helpful discussions. The project was supported by the Deutsche
Forschungsgemeinschaft (DFG) and the Croatian Ministry of Science, Education and
Sports under Grant 035-0000000-2836.

\end{document}